\documentclass[journal=jpclcd,manuscript=article]{achemso}
\usepackage[version=3]{mhchem} 
\usepackage[T1]{fontenc}       
\usepackage{subfigure}

\author{Martin Streiter}
\affiliation{Institut f\"{u}r Physik, Technische Universit\"{a}t Chemnitz, 09126 Chemnitz, Germany}
\altaffiliation{These authors contributed equally to this work}
\author{Stefan Krause}
\altaffiliation{These authors contributed equally to this work}
\email{stefan.krause-physik@web.de}
\affiliation{Institut f\"{u}r Physik, Technische Universit\"{a}t Chemnitz, 09126 Chemnitz, Germany}
\author{Christian von Borczyskowski}
\affiliation{Institut f\"{u}r Physik, Technische Universit\"{a}t Chemnitz, 09126 Chemnitz, Germany}
\author{Carsten Deibel}
\affiliation{Institut f\"{u}r Physik, Technische Universit\"{a}t Chemnitz, 09126 Chemnitz, Germany}

\title[Dynamics of Single Molecule Stokes Shifts]{Dynamics of Single Molecule Stokes Shifts: Influence of Conformation and Environment}

\keywords{Single Molecule Spectroscopy, Molecular Dynamics, Excitation Spectroscopy, Stokes Shift}

\begin{document}

\noindent This document is the Accepted Manuscript version of a Published Work that appeared in final form in The Journal of Physical Chemistry Letters, copyright \textcopyright  American Chemical Society after peer review and technical editing by the publisher. To access the final edited and published work see 
DOI: 10.1021/acs.jpclett.6b02102\\
https://pubs.acs.org/articlesonrequest/AOR-JggM5eevqNDv9ddqMPhM

\begin{tocentry}
 \includegraphics*{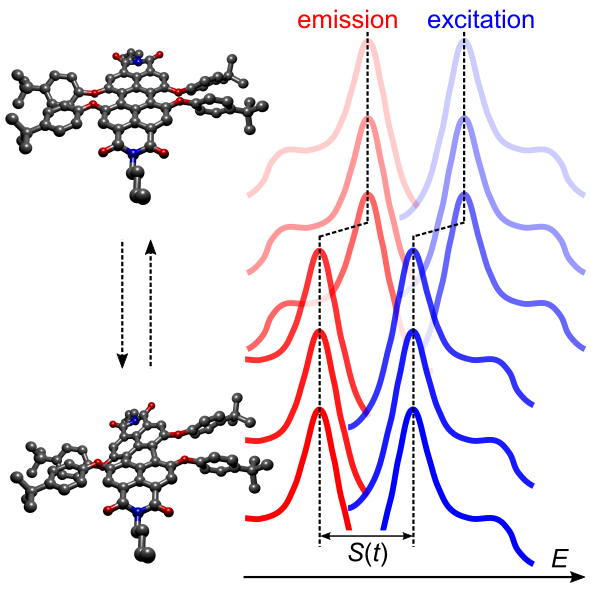}
\end{tocentry}

\begin{abstract}
\noindent We report on time dependent Stokes shift measurements of single molecules. Broadband excitation and emission spectroscopy were applied to study the temporal Stokes shift evolution of single perylene diimide molecules (PDI) embedded in a polymer matrix on the   time scale of seconds. The Stokes shift varied between individual molecules as well as for single molecules undergoing different conformations and geometries. From the distribution and temporal evolution of Stokes shifts, we  unravel the interplay of nano-environment and molecular conformation. We found that Stokes shift fluctuations are related to simultaneous and unidirectional shifts of both emission and excitation spectra. 
\end{abstract}

The first experiments on optical detection of a single molecule were based on resonant excitation at low temperatures. \cite{Moerner1989,Orrit1990}  The excitation spectra measured by the sweep of a narrowband dye laser provided an unexpected wealth of information such as fluorescence lifetimes, \cite{PIROTTA1993} quantum jumps, \cite{Orrit1990} spectral diffusion processes, \cite{Ambrose1991} photon antibunching, \cite{Basche1992} blinking dynamics \cite{Moerner1997} and magnetic resonance effects. \cite{Wrachtrup1993} 
Single molecule spectroscopy  has developed into a standard technique in many aspects of science. \cite{Moerner2002,Kulzer2004,Lu1998,Woll2009}  Extending the applicability of single molecule spectroscopy to room temperatures and living cells supported this progress. \cite{Trautman1994,Xie1996,Xie1998} 

Since these first experiments, the obtainable  information such as emission spectra, photon distributions, positions and orientations of individual emitters have become accessible in the lab as well as by commercial setups. \cite{Krause2011,schmidt2014,krause2016} However, measuring excitation spectra of a single molecule at room temperature remains challenging. Especially for strong phonon broadening,  the red shifted fluorescence signal is weak, increasing the acquisition time for an excitation spectrum. Furthermore, at room temperature, the excited molecule is vulnerable to photo-oxidation. Nevertheless, the Stokes shift of a single molecule  can be determined by measuring emission and excitation spectrum simultaneously  and has been reported recently. \cite{Blum2011,Stopel2014,Piatkowski2016} Stopel \textit{et al.} applied a tuneable laser source to excite single molecules at discrete steps of different wavelengths to reconstruct the excitation spectrum. They found that single molecule Stokes shifts are not inherently constant for chemically equivalent dye molecules. The experimentally determined Stokes shifts formed a broad distribution. \cite{Stopel2014} In the most recent study by Piatkowski \textit{et al.}, excitation spectra were acquired for about two minutes by a broadband Fourier approach. \cite{Piatkowski2016} The method's  advantage is the robustness against blinking and bleaching of the observed molecule. While these studies already revealed unexpected information on the photo-physics of single molecules, the temporal evolution of the Stokes shift during spectral diffusion processes remained unknown. It is worthwhile to push the temporal resolution of excitation and emission acquisition to the  time scale of seconds and below where spectral diffusion can  be observed. This information is not only desirable from a photo-physical point but is of great importance for single molecule F\"orster resonance energy transfer (FRET) based experiments where the spectral overlap between emission of donor and  absorption of acceptor strongly impact the transfer efficiency. \cite{roy2008practical,schuler2008protein,ha1996probing,beljonne2009beyond}

Here we present an experiment  that allows us to study the temporal evolution of emission, excitation and Stokes shift of photo-stable single PDI molecules in a polymer matrix. PDI molecules of this type have been shown to exhibit strong and temperature-dependent fluctuations between different conformations. \cite{Kowerko2009, Krause2011,osswald2005perylene,Fron2008,Vallee2004,Hofkens2001} As a consequence, these molecules show strong shifts in emission energies while the impact on the absorption remains unknown. We explored differences between individual molecules and their time-dependent behavior, revealing the influence of nano-environment and molecular conformation.

\begin{figure}[t]
  \includegraphics*[scale=0.4]{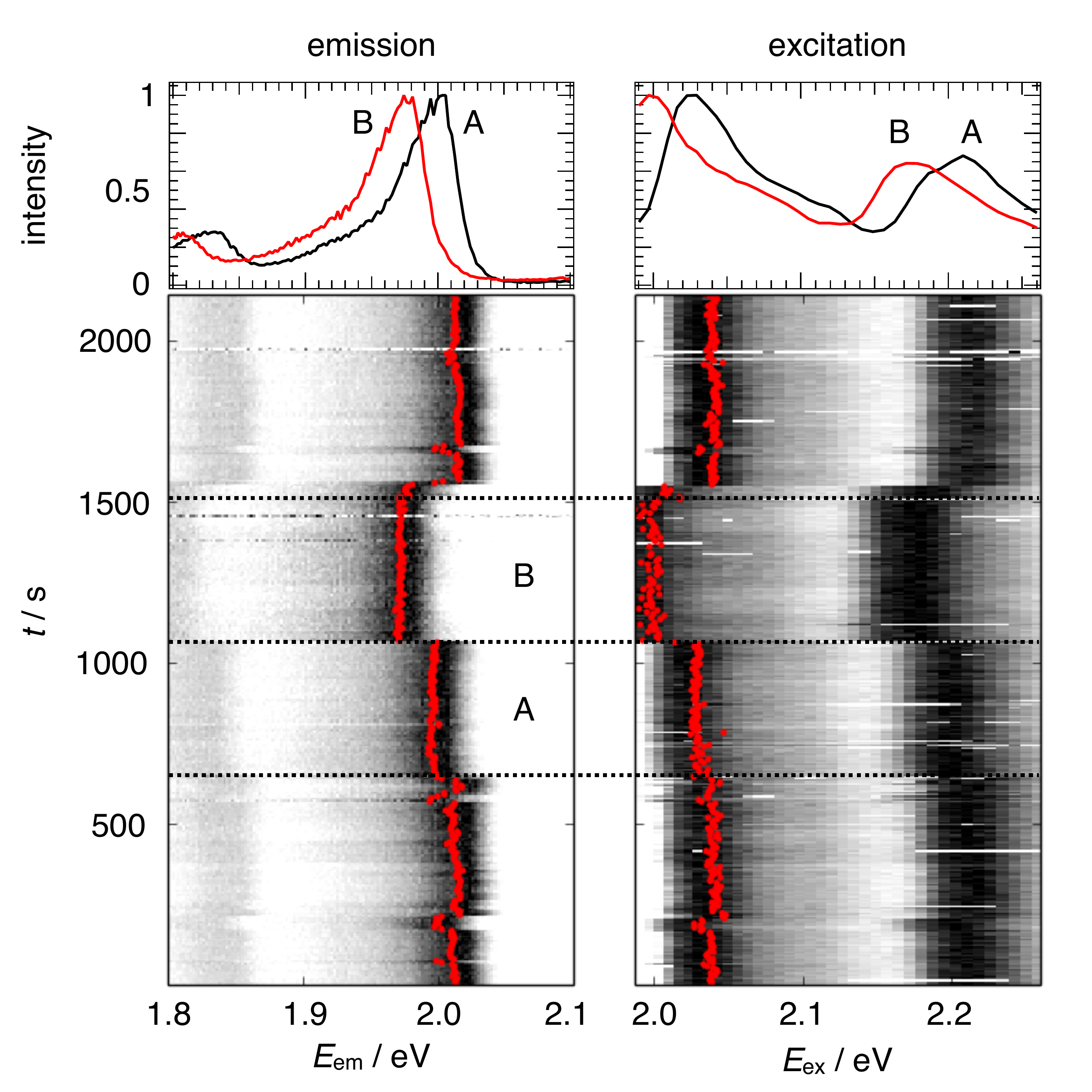}
    \caption{ Time dependent emission and excitation spectra of a single PDI molecule at 77\,K. Red dots indicate the peak positions gained from fitting the spectra. Changes between discrete states of emission and excitation are simultaneously occurring. The averaged and normalized spectra of conformation A and B (marked by dashed lines) are presented above the time-trace.}
    \label{timetrace77}
    \end{figure}

Single PDI molecules were studied at room temperature under ambient conditions and at 77\,K in vacuum by confocal microscopy as explained in the experimental section. At room temperature, 38 single molecule  time-traces  with a length of $t = 10-160\,$s were recorded. At 77\,K, 32 time-traces were measured. The accessible length of time-traces varied depending on the  photo-stability between  $t = 20-2000\,$s. Figure \ref{timetrace77} shows a representative  example of 400 consecutive emission and excitation spectra of a single PDI molecule at 77\,K recorded for 37\,minutes. We determined  the  Stokes shift from  the difference of the main emission and excitation peaks (marked as red dots in the Figure). Both emission and excitation spectra show simultaneous and discrete energetically different  transitions. The investigated PDI molecules are known to form different conformations depending on the orientation of the four substituents in bay positions.\cite{Hofkens2001,Fron2008,Kowerko2009,Krause2011} The chemical structure of  PDI is shown in the experimental section. Transitions between different molecular conformations have been extensively studied and are observable even for molecules embedded in polymer matrices beyond the glass transition temperature.\cite{Hofkens2001,Vallee2004,Kowerko2009,Krause2011} In the present example, several distinct spectral jumps were observed. The emission spectrum shifts over an energy range of about 50\,meV which is comparable to the overall difference in transition energies for all possible conformations of 88\,meV calculated by Fron \textit{et al.} \cite{Fron2008} Additionally, we would like to mention that the energy range of the emission and excitation peak positions covered by all measured molecules in this study is about 100\,meV.

For $t\,=\,630-1050$\,s (state A) and $t\,=\,1050-1520$\,s (state B) in Figure \ref{timetrace77} the averaged emission and excitation spectra are shown above the time-traces. The main emission peak appears at 2.003\,eV for state A and at 1.977\,eV for state B. The corresponding excitation spectrum shows a comparable main peak at 2.036\,eV for state A and 2.000\,eV for state B, which is assigned to transitions from the $S_0$ ground state to $S_1$ states.  Additionally, emission and excitation spectra feature at least one distinct vibronic sideband. The energy differences for two consecutive states A and B, either in emission or excitation, were calculated from the peak energy differences $\Delta E = E_B - E_A$. From emission and excitation maxima  follows a Stokes shift of 33\,meV in state A, which decreases to 23\,meV in state B. This change in Stokes shift of 10\,meV is small in comparison to the change in the peak position of emission $\Delta E_{\mathrm{em}} = 26$\,meV and excitation $\Delta E_{\mathrm{ex}}=36$\,meV. These comparably small changes of the Stokes shift apply for all spectral fluctuations in Figure \ref{timetrace77} and are typical for every measured single molecule.

\begin{figure}[t]
  \includegraphics*[scale=0.7]{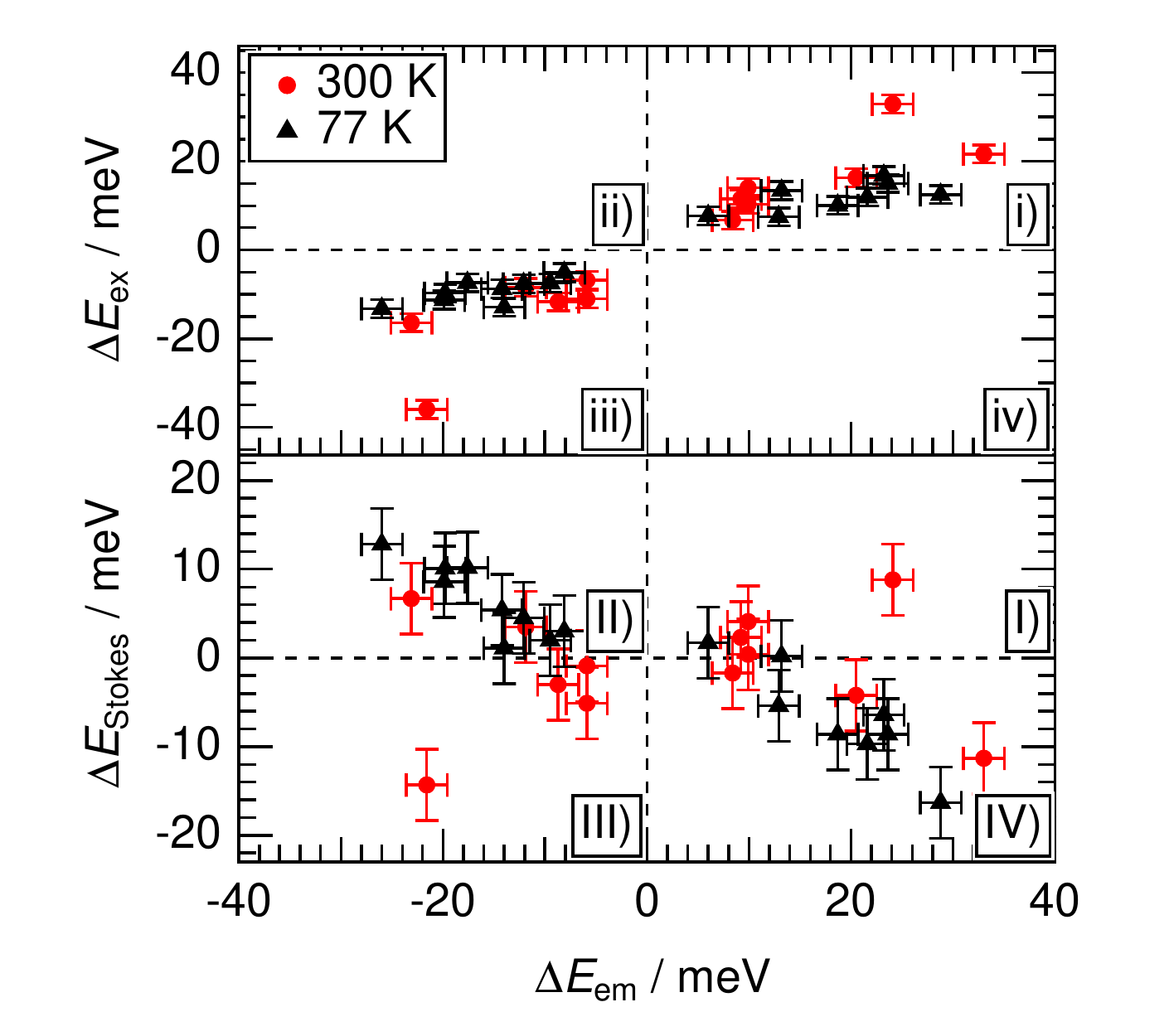}
    \caption{Quadrants i) to iv): spectral jumps  $\Delta E_{\mathrm{ex}}$ vs. $\Delta E_{\mathrm{em}}$  above a threshold of 5\,meV at 300\,K (red) and 77\,K (black). Quadrants I) to IV):  Stokes shift variations $\Delta E_{\mathrm{Stokes}}$  for the spectral jumps  above.}
    \label{delta1}
    \end{figure}

For a detailed discussion of how conformational changes  influence  emission and excitation properties, we evaluated  for all molecules every spectral jump above a threshold of 5\,meV.   Smaller spectral fluctuations were  observed but neglected due to their high fluctuation frequency and therefore insufficient evaluation accuracy. We observed only minor changes of spectral shape and intensity during these jumps.   Spectral jumps were found to occur simultaneously within our experimental time resolution (3\,s at 300\,K and 5.5\,s at 77\,K for one emission and excitation spectrum).    Figure \ref{delta1} (top) shows a nearly linear and temperature-independent  correlation between $\Delta E_{\mathrm{em}}$  and $\Delta E_{\mathrm{ex}}$.  It can be seen that emission and excitation always shift synchronously to higher or lower energies, respectively (quadrants  i and iii). No anti-correlation or opposite dynamics could be observed  (quadrants  ii and iv). 
 
We further examined the variations of the Stokes shift $\Delta E_{\mathrm{Stokes}}$ as a function of emission energy fluctuations $\Delta E_{\mathrm{em}}$ in Figure \ref{delta1} (bottom). At 300\,K, increasing and decreasing Stokes shifts were measured during spectral fluctuations. They were caused by different jump widths of emission or excitation, respectively. These processes are reversible. Consequently, spectral jumps appear in quadrants I to IV.  At 77\,K,  spectral jumps appear only in  quadrant II and IV. This means that   shifts of emission energy are larger than shifts of excitation energy at low temperatures. We assume a temperature dependent correlation between emission and excitation shifts. For both temperatures, the largest    Stokes shift changes were $\Delta E_{\mathrm{Stokes}} = \pm 15\,$meV.
  
\begin{figure}[t]
 \subfigure{ \includegraphics*[scale=0.5]{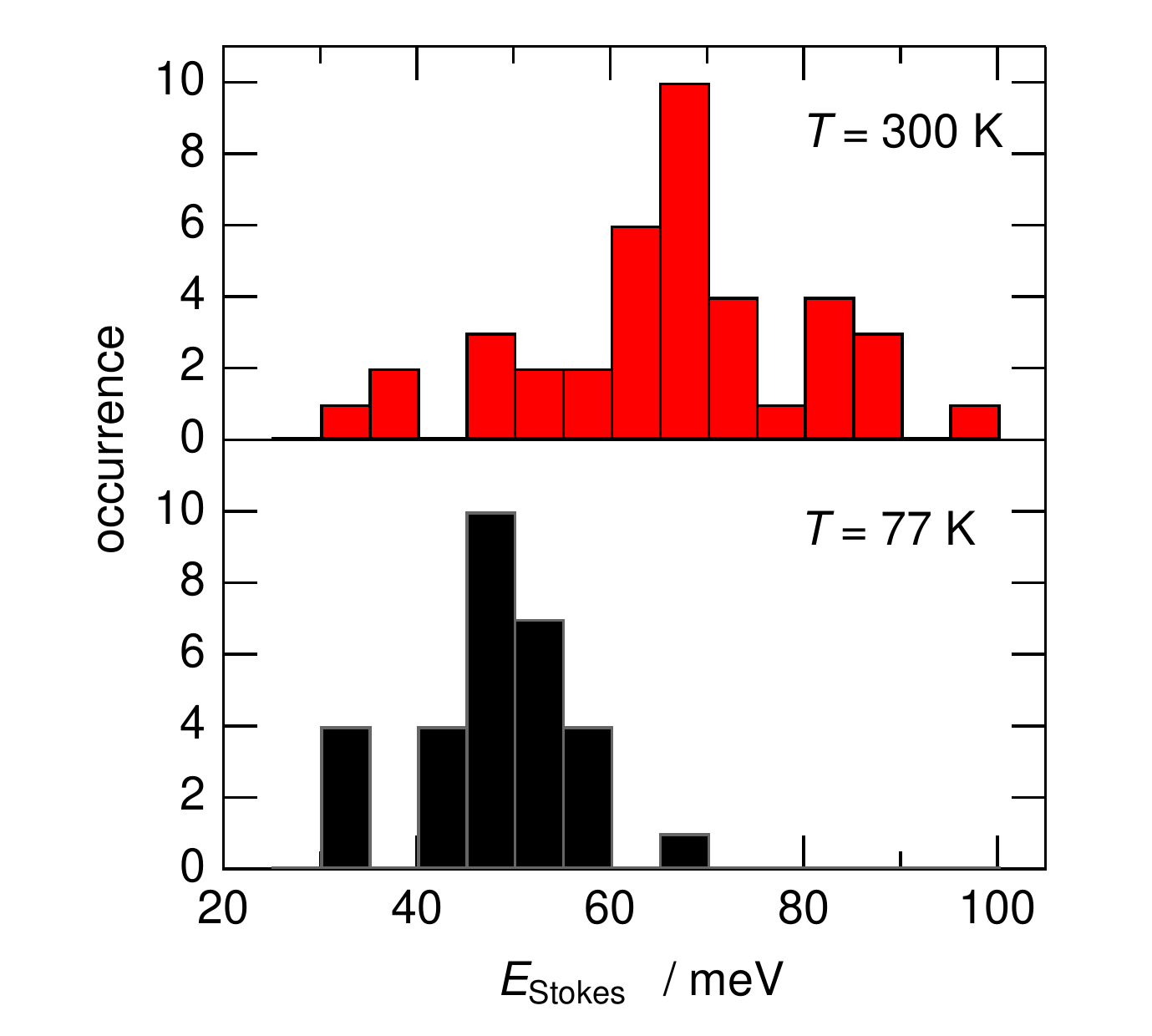}
  \includegraphics*[scale=0.5]{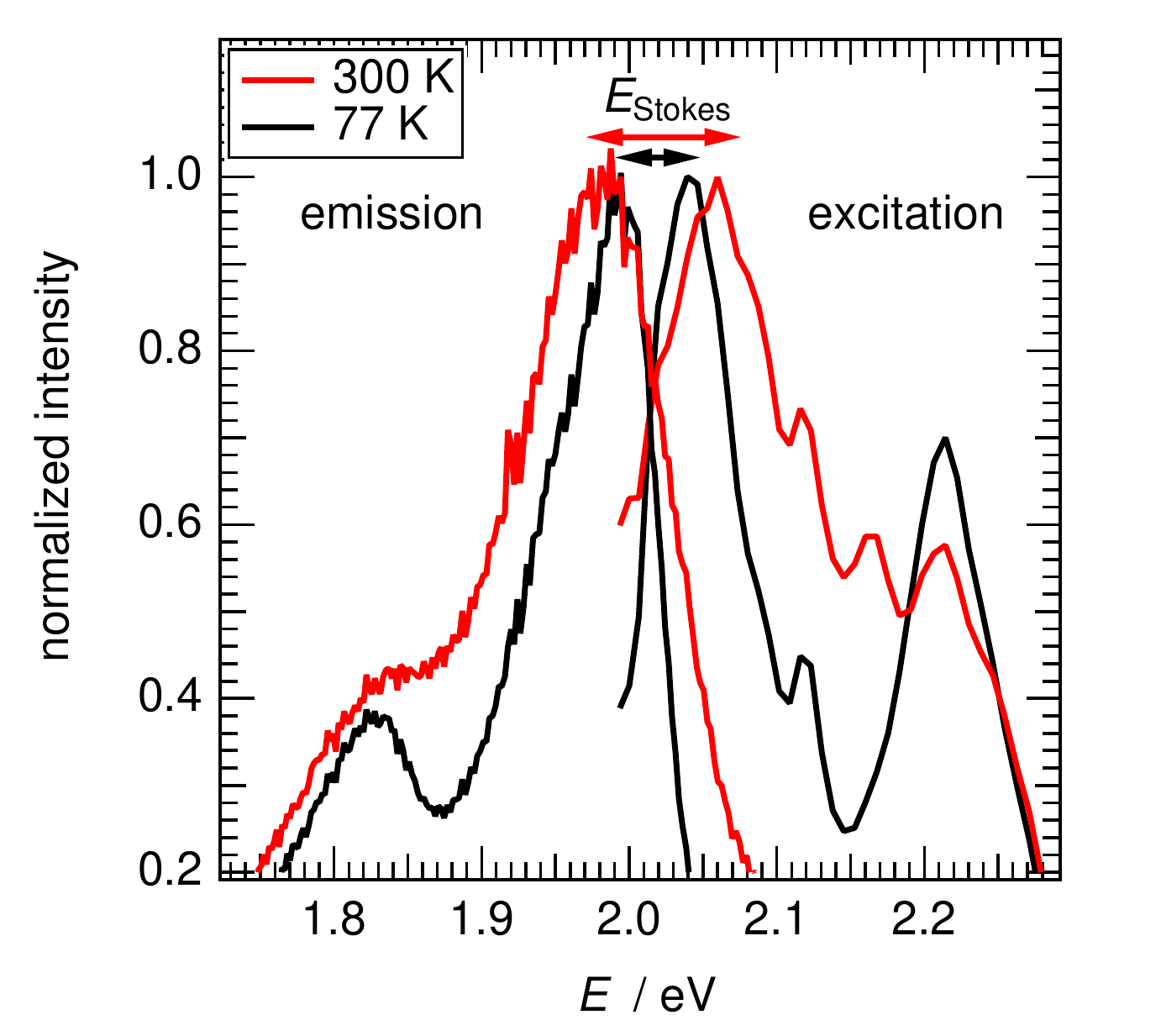} }   
  \caption{Occurrence of average Stokes shift values $\bar E_{\mathrm{Stokes}}$ of single molecule spectral time-traces at 300\,K and 77\,K. With higher temperature, the distribution of Stokes shifts is broader and located at higher values. The right graph shows the average of all spectra measured. The average Stokes shift increases due to  increased line-broadening (indicated by arrows) at 300\,K.}
    \label{histogram}
    \end{figure}

Now  we discuss the  Stokes shift variations $\Delta E_{\mathrm{Stokes}}$ of individual single molecule time-traces with respect to the overall  Stokes shift   distribution of all  molecules measured.  Figure \ref{histogram} shows the distribution of  average Stokes shifts. Each value results from  calculating the mean Stokes shift of a complete time-trace. At 300\,K we found a broad distribution of Stokes shifts  centered around 65\,meV. The width of the  distribution shows that Stokes shift  variations of $\Delta E_{\mathrm{Stokes}}\approx 70$\,meV are possible despite chemically identical molecules. This value  far exceeds the maximal Stokes shift variations of 15\,meV which have been found within  individual spectral time-traces. From the broad and temperature-dependent  Stokes shift distribution, we conclude that the nanoscopic environment of the molecules is responsible for large differences in the Stokes shift. This  can be explained by a different strength in interaction of individual molecules either to their own phenoxy side groups \cite{Fron2008} or  the surrounding polystyrene matrix. Large spectral jumps in emission and excitation  up to 80\,meV as observed for individual molecules within a spectral time-trace do not influence the Stokes shift to the same extent. Vall\'ee \textit{et al.} have shown that geometrical fluctuations  can occur within the  volume available for geometrical rearrangements of the molecule in a polymer below the glass transition temperature. \cite{Vallee2004} These changes do not require a large rearrangement of the chromophoric PDI backbone with respect to the polymer environment. We suggest that the resulting impact on coupling is comparably small, leading to minor Stokes shift variations during measurement of an individual molecule.  This is reasonable since the phenoxy side groups are immobilized by the rigid polymer below the  glass transition temperature. Temperature lowering to 77\,K results in a narrowed distribution of Stokes shifts centered at 50\,meV. The smaller Stokes shifts at 77\,K can be assigned to a  reduction of the phonon wing by annihilation of phonons at low temperatures. \cite{Friedrich1984} The phonon wing is responsible for the line-broadening of the main emission and excitation peak at elevated temperatures resulting in a dislocation of peak centers and therefore  larger Stokes shifts as shown in Figure \ref{histogram}.
 
In conclusion, we performed time dependent emission and broadband excitation spectroscopy of single perylene diimide molecules on the time scale of seconds. Varying states of emission and excitation energies were observed and attributed to different fluctuating molecular geometries caused by interactions with the heterogeneous environment. We found that  emission and excitation spectra change at the same time. Emission and excitation spectra  showed simultaneous red or blue shifts, respectively. Spectral jump widths of emission and excitation  were found to be different, leading to Stokes shift variations with a maximum difference of 15\,meV within single molecule spectral time-traces. Since the molecules are immobilized within the polymer matrix below glass transition temperature, we assume that these minor variations are caused by  twisting of the chromophoric PDI backbone.  \cite{Vallee2004} In contrast, stronger variations of the average  Stokes shifts between different molecules were observed,  covering an energy range of 70\,meV.  This can be explained by  different coupling of  the static nanoscopic environment and the phenoxy side groups to the PDI chromophoric backbone. 

\section{Experimental}

\begin{figure}[t]
\includegraphics*[scale=0.5]{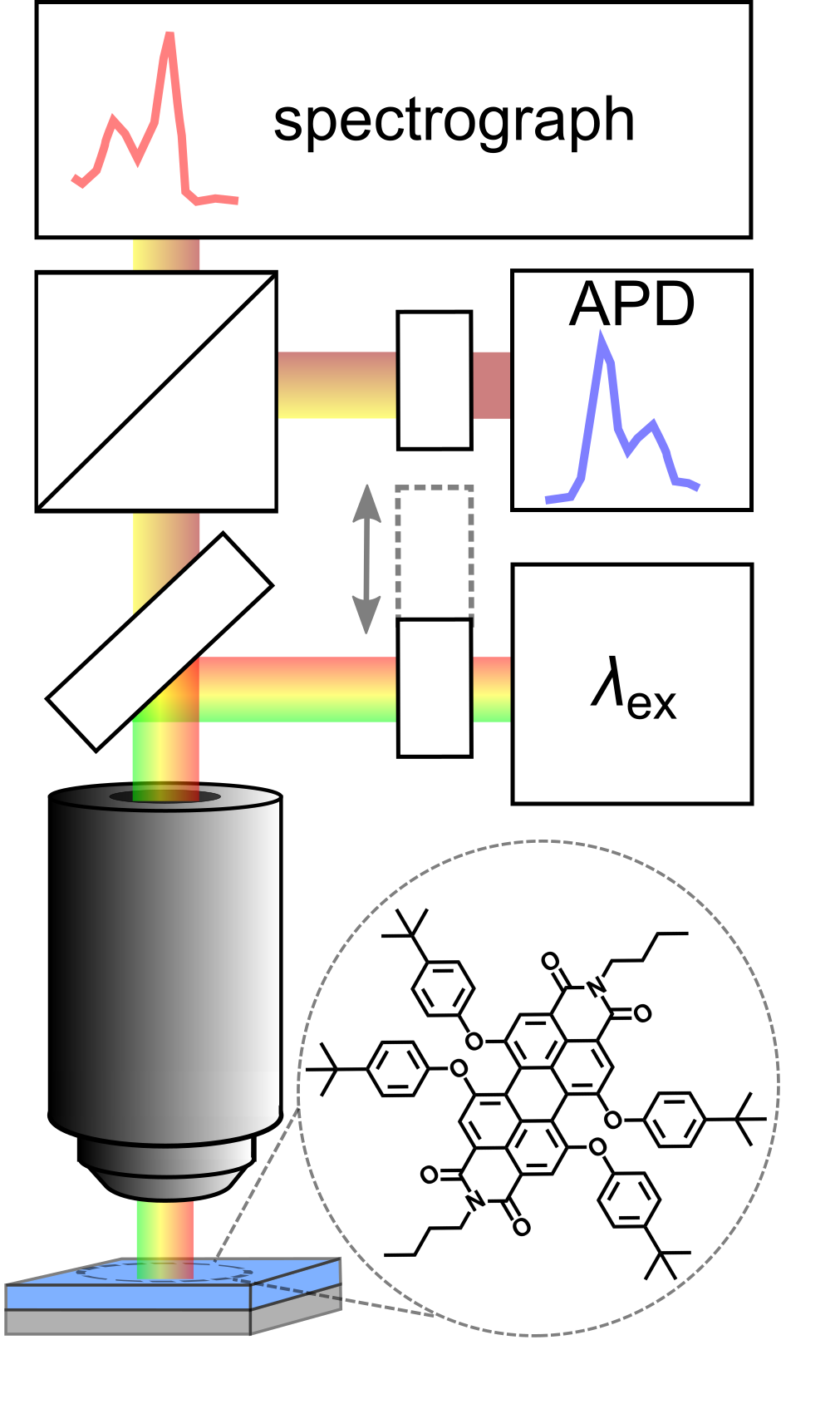}
  \caption{Scheme of the experimental setup. A  supercontinuum laser provides a tuneable excitation  wavelength  $\lambda_{\mathrm{ex}}$. In combination with achromatic beam splitters,  fast switching between emission (spectrograph) and excitation acquisition (ADP) was realized.   }
    \label{setup}
    \end{figure}

The perylene diimide derivate N,N'-Dibutyl-1,6,7,12 - tetra (4- tert-butylphenoxy) perylene-3,4:9,10-tetracarboxylic acid bisimide (PDI) was synthesized by W\"{u}rthner \textit{et al.}\cite{wurthner2001fluorescent} and embedded in a poly(styrene) (PS) matrix with a molecular weight of $20\,\mathrm{kg/mol}$. A PS toluene solution was doped with $10^{-10}\,\mathrm{mol/l}$ PDI and spincoated onto silicon dioxide substrates. Single molecule confocal microscopy and emission spectroscopy as a function of obesrvation time were performed as previously described. \cite{Krause2011a} For excitation spectra acquisition, a tuneable supercontinuum white light laser source (SuperK EXR-15 with SELECTplus filter,  NKT Photonics) was used. Excitation spectra were measured by detecting the photoluminescence above 640\,nm with a longpass filter. Emission spectra were measured with an excitation wavelength of $\lambda_{\mathrm{ex}} = 560\,\mathrm{nm}$.  A non-polarizing 50:50 beam splitter separated the fluorescence signal into two equal fractions for  subsequent excitation and emission measurements. A scheme of the experimental setup is given in Figure \ref{setup}. At 300\,K, emission spectra were acquired for 0.5\,s (1\,s at 77\,K) and excitation spectra for 2.5\,s (4.5\,s at 77\,K, longer acquisition times due to cryostate and objective with lower numerical aperture).  Fast switching (10\,ms) between  acquisition of emission and excitation  was realized by utilizing an  automated   bandpass filter and a spectrograph shutter.

\section{Author Information}

\subsection{Corresponding Author}
Email: stefan.krause-physik@web.de

\subsection{Notes}
The authors declare no competing financial interest.

\begin{acknowledgement}
This work was partly funded by the DFG (project DFG DE830/13-1). We thank Frank W\"urthner (Julius-Maximilian-University W\"urzburg) for a kind donation of PDI.  
\end{acknowledgement}

\bibliography{bibliography}

\end{document}